\documentclass[aps,prl,twocolumn,amsmath,amssymb,floatfix,superscriptaddress,showpacs]{revtex4}
\usepackage{graphicx}

\begin{document}


\title{3D magnetization profile and multi-axes exchange bias in Co antidot arrays}



\author{F.~Fettar}
\author{L.~Cagnon}
\author{N.~Rougemaille}
\affiliation{Institut N\'eel, CNRS $\&$ Universit\'e Joseph Fourier, BP166, F-38042 Grenoble Cedex 9, France}
%


\date{\today}

\begin{abstract}
Cu/Co/Cu trilayers have been deposited on nanoporous alumina membranes. Magnetic properties of the resulting Co antidot arrays are investigated using SQUID magnetometry. Hysteresis loops of these arrays show two-step magnetization reversal. In addition, exchange bias is observed, whether the cooling field is applied within or perpendicular to the surface plane. In the former case, the exchange bias changes sign close to the blocking temperature, and becomes positive. We attribute these effects to the local, crescent shape of the Co films, induced by the surface morphology of the alumina membranes. This morphology leads to a three-dimensional magnetization distribution at the nanoscale. 
\end{abstract}

\maketitle


Advances in lithography techniques, material engineering and elaboration processes allow fabrication of structures with size and geometry controlled at the nanoscale. These  developments stimulate research in nanotechnology and offer new opportunities in nanosciences. This is particularly true for nanomagnetism: systems with physical dimensions comparable to, or smaller than, the characteristic length scales involved in magnetism are now widely accessible experimentally \cite{Bader06,Nogues05,Skomski03}. Magnetic antidot arrays, i.e. magnetic thin films with a periodic array of nonmagnetic inclusions, are one example of system in which magnetic properties can change in various ways. If antidot arrays are exciting scientific playgrounds for fundamental research \cite{Heyderman06,Liu01,Tripathy08,Wang05,Yu03}, they are also promising candidates for high density information storage \cite{Cowburn97,Torres98,Rahman08}. For these applications, the ability to control the strength and orientation of magnetic anisotropy is essential, especially for the thermal stability and switching reliability of magnetic bits. Exchange biasing is another important ingredient that often plays a crucial role in devices. Most of the time, these two key magnetic properties are optimized in multi-materials, patterned architectures.

In this work, we show that Co films, deposited on nanoporous alumina membranes used as templates to fabricate highly-ordered ferromagnetic antidot arrays \cite{Xiao02,Rahman07,Navas07}, exhibit several unexpected features: i) in-plane hysteresis loops evidence a two-step magnetization reversal process, ii) remanence is found for both in-plane and out-of-plane components of magnetization, iii) after a field cooling procedure, exchange bias is observed for in-plane and out-of-plane applied magnetic fields, and iv) close to the blocking temperature, the exchange bias induced by an in-plane cooling field changes sign, and becomes positive. These effects are intimately related to the surface morphology of the alumina membranes that leads locally, at the nanoscale, to a three-dimensional magnetization distribution in the Co film.

Cu(10 nm) / Co(8 nm) / Cu(10 nm) trilayers were dc magnetron-sputtered at room temperature onto nanoporous alumina membranes produced using electrochemical oxidation of pure aluminum \cite{Masuda95}. These membranes are obtained after a two-step anodizing procedure, dissolution of the aluminum substrate, chemical etching of the barrier layer, and pore widening treatment in a diluted phosphoric acid solution. Scanning electron microscopy (SEM) reveals that mono-disperse nanopores self-organize during the anodizing process and are located on a hexagonal network. Here, pore diameter and interpore distance are about 60 and 105 nm, respectively. Figure 1 shows a typical 11x7 $\mu$m$^2$ SEM image after deposition of the trilayer: no significant reduction of the pore diameter is found after completion of the $\sim$30 nm-thick film. Magnetic properties of these antidot arrays have been investigated using SQUID (superconducting quantum interference device) magnetometry.

\begin{figure}
\includegraphics[width=8.5cm]{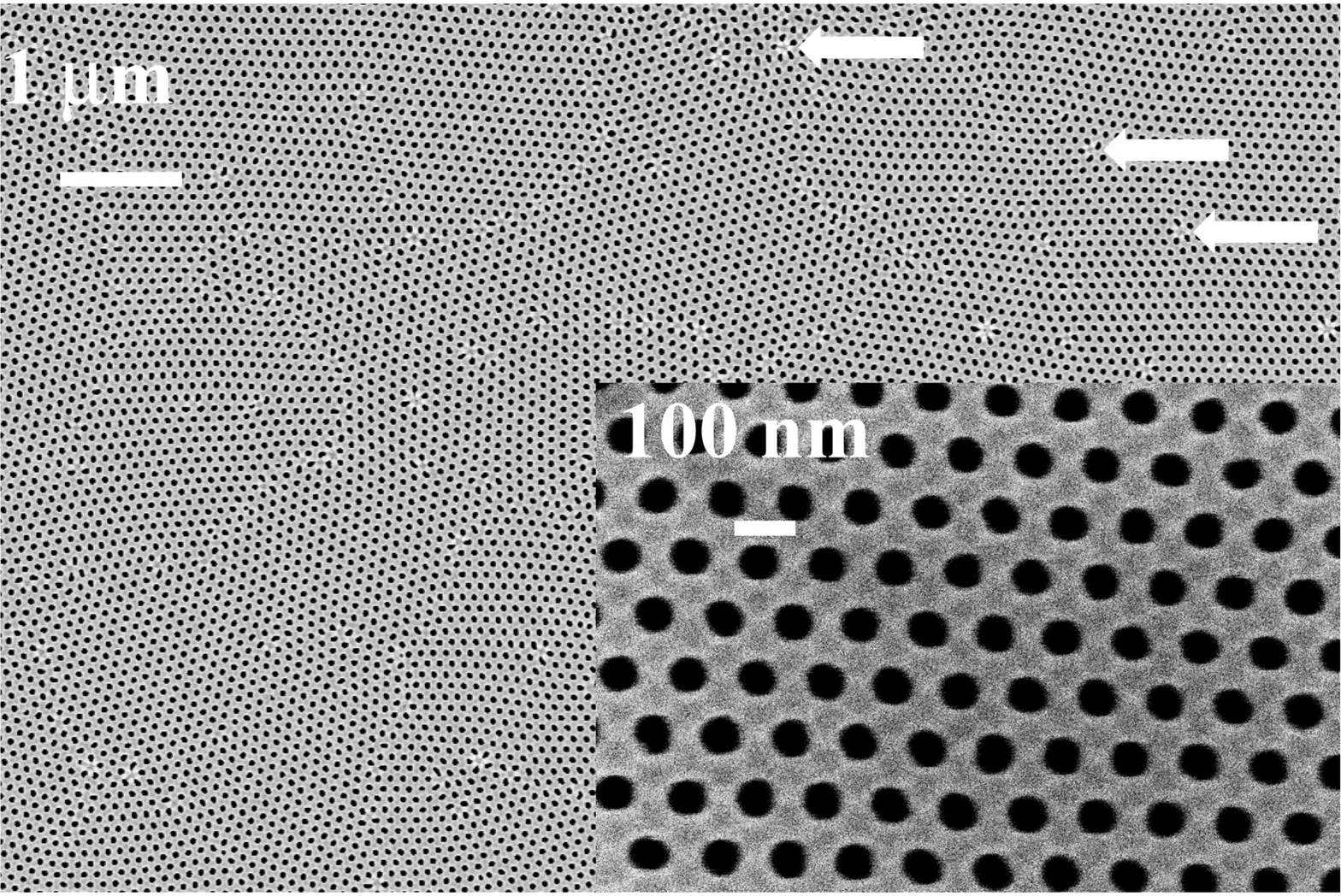}
\caption{\label{fig1}
SEM image of a Cu(10 nm)/Co(8 nm)/Cu(10 nm) trilayer deposited on top of a nanoporous alumina membrane. White arrows indicate positions of edge dislocations in the hexagonal arrangement of the nanopores. The lower-right inset shows a higher magnification image of the antidot array.
}
\end{figure}

Surprisingly, the in-plane hysteresis loop measured at room temperature shows a two-step magnetization reversal (Fig. 2a). This is puzzling at first sight since only one ferromagnetic film is present in the trilayer. We might think that the hexagonal symmetry of the alumina membranes plays an important role in the reversal process. However, angular measurements do not reveal significant in-plane anisotropy: the shape of the loops is found to be independent of the field orientation, and we only observe weak variations of the coercivity and remanence ($\sim$10\% and $\sim$3\% respectively) when rotating the applied magnetic field in the film plane \cite{note1}. This is consistent with the fact that nanopores are perfectly ordered within grains that extend over several $\mu$m$^2$, but which are misoriented on a larger scale (see edge dislocation indicated by white arrows in Fig. 1). Thus, although the system locally has threefold symmetry, this symmetry is lost macroscopically, and no in-plane anisotropy is expected. Another explanation to interpret the two-step reversal process could be that the Co deposited inside the nanopores contributes to a magnetic signal. However, the nanopores have an aspect ratio of about 1000 (60 nm in diameter, $\sim$60 $\mu$m long), and it is very unlikely that this additional Co can coat the inner walls of the nanopores and give rise to a strong ferromagnetic contribution at room temperature. A more plausible scenario is that the Co film has both in-plane and out-of-plane easy axes. Consistent with this interpretation, we find a clear remanent out-of-plane component of magnetization when the field is applied perpendicular to the surface plane (Fig. 2b). Presumably, this unusual magnetization distribution is related to the surface morphology of the alumina membrane. In fact, films deposited on top of these templates are known to have crescent shape \cite{Rahman08}, as sketched in Fig. 2c. We thus attribute the in-plane two-step reversal process to a three dimensional profile of the Co magnetization: magnetic moments between nanopores are aligned within the film plane, while moments along the walls of the nanopores are aligned perpendicular to the surface (Fig. 2c), due to a strong shape anisotropy. The local morphology of the Co film is then similar to what is observed when ferromagnets are sputtered on top of polystyrene or silica nanospheres \cite{Albrecht05,Eimüller08}.

\begin{figure}
\includegraphics[width=8.5cm]{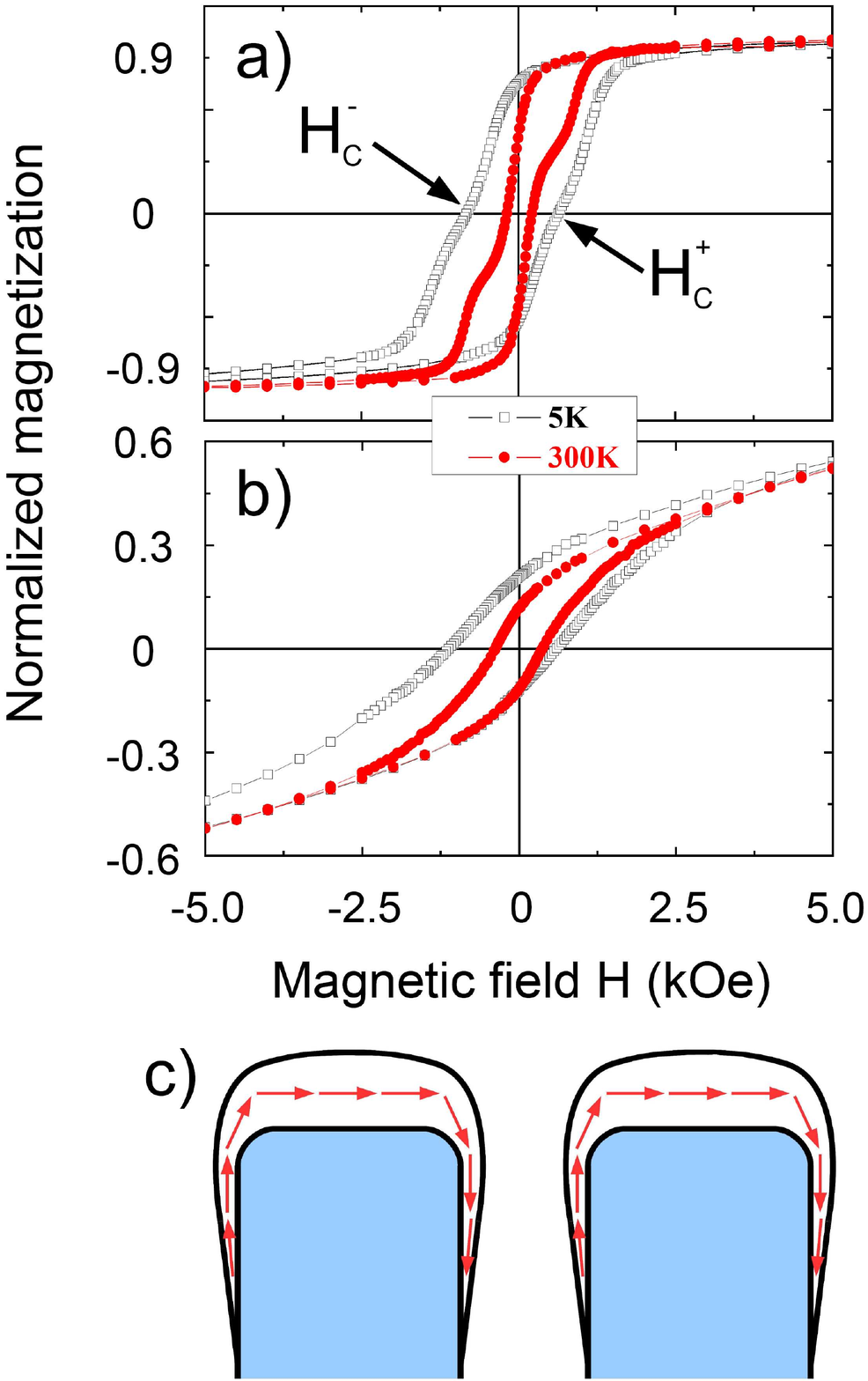}
\caption{\label{fig2}
(Color online) Hysteresis loops at 5 and 300 K for planar (a) and perpendicular (b) measurements. The saturation magnetization for in-plane measurements is used to normalize the SQUID signals. Low temperature loops are obtained after a field-cooling procedure (+1 T applied magnetic field). (c) Cross-section schematics of the local morphology of the Co film. 
}
\end{figure}

Then, the trilayers have been cooled down from room temperature to 5 K under a 1 T magnetic field, applied either within the film plane or perpendicular to the sample surface. In the following, all the hysteresis loops were measured between $\pm$2 T for an applied external magnetic field \textit{parallel} to the cooling field. Measurements clearly reveal exchange bias at low temperature in both cases. Since there is no intentional deposition of an antiferromagnetic material, this exchange bias effect is attributed to a partial oxidation of the Co layer when the sample is exposed to air, and the formation of CoO. This could be surprising as the Cu capping layer is 10 nm thick, and should prevent the ferromagnetic layer from oxidation \cite{note2}. However, the oxidation process could be preferentially favored in the vicinity of the antidot, where the overall thickness of the trilayer is reduced due to the crescent shape of the deposit. In other words, CoO rings could form along the walls of the nanopores, similarly to what is observed in Co nanopillars and nanohills \cite{Balcells09,Rosa09}. This scenario implies that the Co layer is essentially oxidized where magnetization points out of the surface plane, while regions where magnetization is purely in-plane remain metallic. Consistent with this picture, we find that the exchange bias effect is significantly stronger when the antidot array is cooled down in a magnetic field perpendicular to the sample surface. Estimation of the exchange field $H_{E}$ (average value of the positive ($H_{C}^{+}$) and negative ($H_{C}^{-}$) coercive fields shown in Fig. 2a) at 5 K for in-plane and out-of-plane configurations gives -105 Oe and -262 Oe, respectively. This difference is even more striking when compared to the ratio $r$ of the in-plane / out-of-plane remanence ($r$ $\sim$ 4) measured at 5 K. Although the density of in-plane magnetic moments is several times higher than the density of out-of-plane moments, $H_{E}$ is 2.5 larger in the latter case.

Finally, the ferromagnetic antidot arrays have been warmed up from 5 K to room temperature, and hysteresis loops were recorded at several intermediate temperatures. Variations of the coercive and exchange fields for in-plane and out-of-plane geometries are reported in Figure 3. For in-plane measurements, the exchange bias clearly changes sign close to the blocking temperature.
Positive exchange bias is usually observed in systems with antiferromagnetic interfacial exchange coupling \cite{Nogues00}. Although this is not the case here \cite{Parker00}, positive bias has been already reported in Co/CoO systems \cite{Gredig02,Radu03}, and two main mechanisms could apply to our samples: i) a distribution of blocking temperatures in the CoO grains of our polycrystalline films \cite{Gredig02}, and/or ii) an antiferromagnetic superexchange coupling at the Co/CoO interface originating from roughness or thickness variations \cite{Radu03}. Close to the blocking temperature, we do not observe a non-monotonic behavior of the positive coercive field $H_{C}^{+}$, as expected from Ref. \onlinecite{Gredig02}, but the temperature dependence of $H_{C}^{+}$ and $H_{C}^{-}$ are similar to those reported in Ref. \onlinecite{Radu03}. We thus conclude that the second scenario is more plausible. 

\begin{figure}
\includegraphics[width=8.5cm]{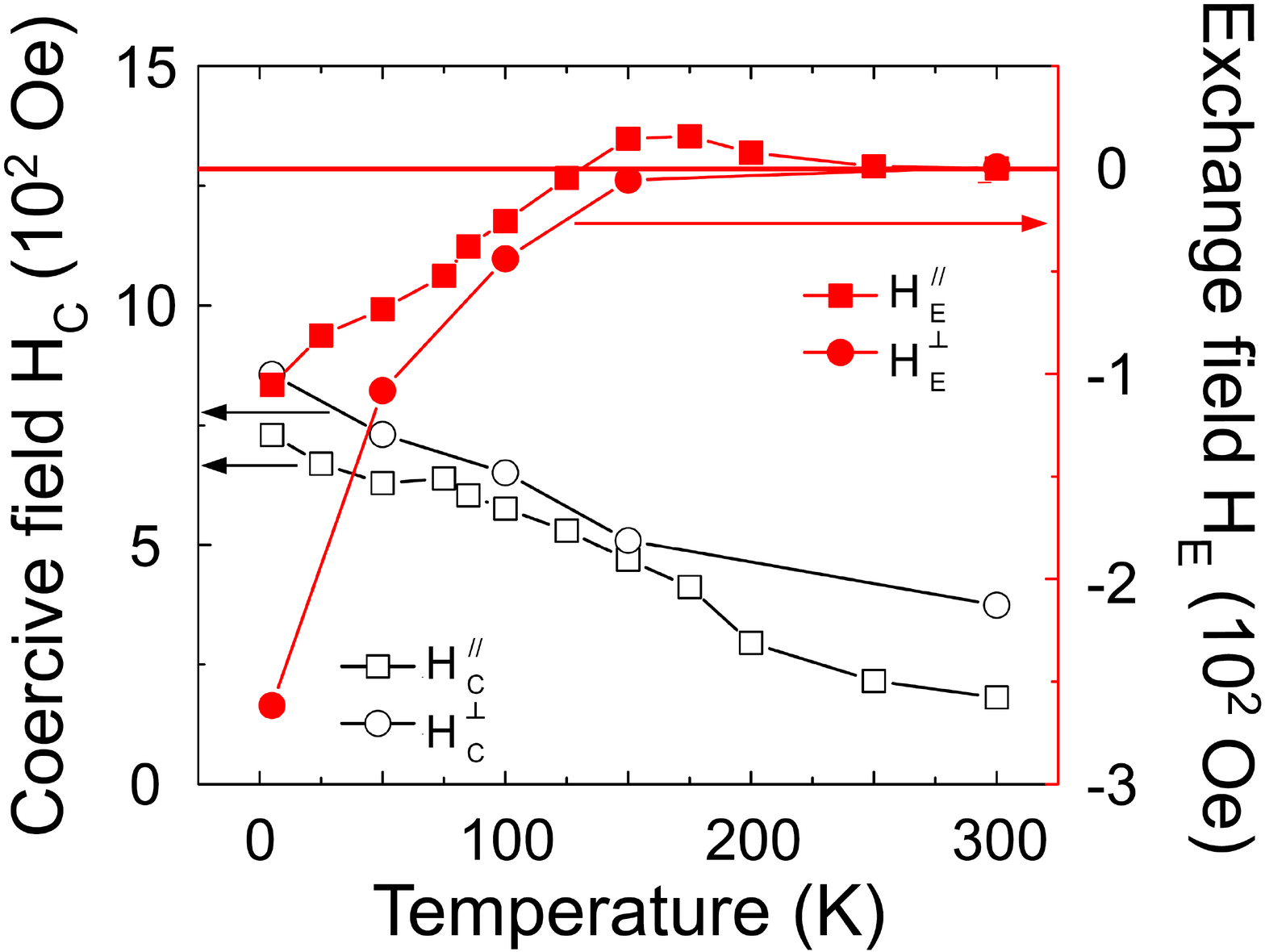}
\caption{\label{fig3}
(Color online) Temperature dependence of the coercive field (open symbols) and exchange field (full symbols), for in-plane (squares) and out-of-plane (circles) measurements. Lines are only guide to the eyes.
}
\end{figure}

In summary, nanometer-thick Co antidot arrays prepared on nanoporous alumina membranes show several interesting magnetic properties: a two-step reversal process, a three dimensional magnetization profile, in-plane and out-of-plane exchange bias, and inversion of the sign of this bias for in-plane measurements. These effects are attributed to the morphology of the alumina template that favors a crescent-shape geometry of the deposit at the vicinity of the nanopores. Magnetic properties of Co antidot arrays may open new opportunities for data storage applications: the use of nanoporous alumina templates is a simple and cheap bottom-up approach to produce magnetic films with tunable anisotropy and exchange bias. Optimization of the \textit{strength} and \textit{directions} of these two key properties could be achieved by a proper choice of the Co thickness and an intentional oxidation procedure. The pore diameter and the interpore distance are also two degrees of freedom that could be finely adjusted to obtain sizable effects.

The authors thank A. Doupal and M. Darques for fruitful discussions, as well as D. Barral, E. Eyraud and S. Pairis for technical support. We also thanks O. Fruchart and J. Vogel for their critical reading of the manuscript.


\begin{thebibliography}{2}

\bibitem{Bader06}
S.D. Bader, Rev. Mod. Phys. {\bf 78}, 1 (2006).

\bibitem{Nogues05}
J. Nogu\'es et al., Phys. Rep. {\bf 422}, 65 (2005).

\bibitem{Skomski03}
R. Skomski, J. Phys.: Condens. Matter {\bf 15}, R841 (2003).

\bibitem{Heyderman06}
L. J. Heyderman et al., Phys. Rev. B {\bf 73}, 214429 (2006).

\bibitem{Liu01}
K. Liu et al., Phys. Rev.B {\bf 63}, 060403(R) (2001).

\bibitem{Tripathy08}
D. Tripathy, A.O. Adeyeye, and N. Singh, Appl. Phys. Lett. {\bf 93}, 022502 (2008).

\bibitem{Wang05}
C. C. Wang et al., Phys. Rev. B {\bf 72}, 174426 (2005).

\bibitem{Yu03}
C. Yu, M.J. Pechan, and G.J. Mankey, Appl. Phys. Lett. {\bf 83}, 3948 (2003).










\bibitem{Cowburn97}
R. P. Cowburn, A. O. Adeyeye, and J. A. C. Bland, Appl. Phys. Lett. {\bf 70}, 2309 (1997).

\bibitem{Torres98}
L. Torres, L. Lopez-Diaz, and J. I$\tilde{n}$iguez, Appl. Phys. Lett. {\bf 73}, 3766 (1998).

\bibitem{Rahman08}
M.T. Rahman, N.N. Shams, and C.-H. Lai, Nanotechnology {\bf 19}, 325302  (2008).

\bibitem{Xiao02}
Z.L. Xiao et al., Appl. Phys. Lett. {\bf 81}, 2869 (2002).

\bibitem{Rahman07}
M.T. Rahman et al., Appl. Phys. Lett. {\bf 91}, 132505 (2007).

\bibitem{Navas07}
D. Navas et al., Appl. Phys. Lett. {\bf 90}, 192501 (2007).





\bibitem{Masuda95}
H. Masuda and K. Fukuda, Science {\bf 268}, 1466 (1995).

\bibitem{note1}
The small differences we observe probably come from a uniaxial anisotropy induced during deposition. 




\bibitem{Albrecht05}
M. Albrecht et al., Nature Mater. {\bf 4}, 203 (2005).

\bibitem{Eimüller08}
T. Eim$\ddot{u}$ller et al., Phys. Rev. B {\bf 77}, 134415 (2008).

\bibitem{note2}
We emphasize that the formation of CoO must occur ex situ, before the sample is introduced into the SQUID magnetometer. We exclude possible in situ oxidation of the layer, for example due to a residual presence of oxygen in the chamber, since trilayers have been deposited simultaneously on nanoporous alumina membranes and silicon substrates, and do not show exchange bias in the latter case. 

\bibitem{Balcells09}
Ll. Balcells et al., Appl. Phys. Lett. {\bf 94}, 062502 (2009).

\bibitem{Rosa09}
W. O. Rosa et al., J. Appl. Phys. {\bf 106}, 103906 (2009).

\bibitem{Nogues00}
J. Nogu\'es, C. Leighton, and I.K. Schuller, Phys. Rev. B {\bf 61}, 1315 (2000).

\bibitem{Parker00}
F. T. Parker, K. Takano, and A. E. Berkowitz, Phys. Rev. B {\bf 61}, 866(R) (2000).

\bibitem{Radu03}
F. Radu et al., Phys. Rev. B {\bf 67}, 134409 (2003).

\bibitem{Gredig02}
T. Gredig et al., Appl. Phys. Lett. {\bf 81}, 1270 (2002).


%
%
%





\end{thebibliography}
\end{document}